\documentclass[aps,prc,preprint,tightenlines,groupedaddress,nofootinbib,
showpacs,preprintnumbers,amsmath,amssymb,superscriptaddress]{revtex4}
\usepackage{graphicx}
\usepackage{dcolumn}
\usepackage{mathrsfs}
\usepackage{bm}

\begin{document}

\title{Validating relativistic models of nuclear structure against 
       theoretical, experimental, and observational constraints} 
\author{J. Piekarewicz}
\affiliation{Department of Physics, Florida State 
             University, Tallahassee, FL 32306}
\date{\today} 

\begin{abstract}
Relativistic mean-field models of nuclear structure have been
enormously successful at reproducing ground-state properties 
of finite nuclei throughout the periodic table using a handful 
of accurately calibrated parameters. In this contribution we 
use powerful theoretical, experimental, and observational 
constraints --- not employed in the calibration procedure --- 
to validate two such models: NL3 and FSUGold. 
It is observed that FSUGold is consistent with all these 
constraints, except perhaps for a high density equation of state 
that appears mildly softer than required by astronomical 
observations. It is argued that incorporating such constrains 
goes a long way in removing much of the ambiguity left over 
from the standard calibrating procedure.
\end{abstract}
\maketitle 

\section{Introduction}
\label{Introduction}

Mean field descriptions of the ground-state properties of medium to
heavy nuclei have enjoyed enormous success. These highly economical
descriptions encode a great amount of physics in a handful of model
parameters that are calibrated to a few ground-state properties of a
representative set of medium to heavy nuclei. An example of such a
successful paradigm is the relativistic NL3 parameter set of
Lalazissis, Ring, and
collaborators~\cite{Lalazissis:1996rd,Lalazissis:1999}.

Yet by their mere nature, such mean-field models are untested away
from their narrow window of applicability. Whereas models fitted to
the ground-state properties of finite nuclei all tend to agree on the
saturation properties of symmetric nuclear matter, they widely
disagree on its density and isospin dependence~\cite{Brown:2000,
Furnstahl:2001un}. To resolve this ambiguity, an effort has been made
to incorporate into the calibration procedure {\sl breathing-mode}
energies of heavy nuclei with neutron-proton asymmetries
[$b\!\equiv\!(N-Z)/A$] ranging from $b\!=\!0.11$ to
$b\!=\!0.21$~\cite{Todd-Rutel:2005fa, Agrawal:2005ix}. Although these
approaches --- combined with improved experimental data and analyzes
--- have helped narrow down the range of acceptable values of the
incompressibility coefficient of symmetric nuclear matter to
$K\!=\!230\!\pm\!10$~MeV~\cite{Agrawal:2003xb,Colo:2004mj,Garg:2006vc},
the density dependence of both symmetric-nuclear and pure-neutron
matter remains largely undetermined. Thus, it is the aim of this
contribution to test the validity of two accurately-calibrated
relativistic mean-field models --- NL3~\cite{Lalazissis:1996rd,
Lalazissis:1999} and FSUGold~\cite{Todd-Rutel:2005fa} ---
against recent theoretical, experimental, and observational
constraints not employed in the calibration procedure. In this
manner we aim to establish the extent to which models that were 
accurately calibrated around nuclear saturation density may be 
reliably extrapolated to the low- and high-density regimes.

From the theoretical perspective, powerful arguments have provided
critical insights into the behavior of pure neutron matter at low
densities. Indeed, the low-density behavior of dilute Fermi gases with
very large scattering lengths is {\sl universal}, in that its energy
equals that of the free Fermi gas up to a dimensionless universal 
constant of the order of $1/2$~\cite{Carlson:2003,Nishida:2006br}. 
Yet pure neutron matter deviates from unitarity due to its relatively 
large effective range. Fortunately, effective-range corrections to 
unitarity were recently computed by Schwenk and 
Pethick~\cite{Schwenk:2005ka}. Such a model-independent approach 
will be used to test the validity of mean-field models away from 
their region of applicability. Moreover, it will be shown how such 
powerful theoretical constraints may be used to rule out a variety
of accurately calibrated models~\cite{Brown:2000,Furnstahl:2001un}.

Laboratory experiments with heavy ions have played a critical role in
constraining the nuclear equation of state. By tuning the energy of
the colliding beams and the neutron-proton asymmetry, heavy-ion
collisions probe vast regions of the phase diagram. For example, by
compressing nuclear matter to pressures never before attained under
laboratory conditions, the equation of state of {\sl symmetric}
nuclear matter was determined up to densities of about 4-to-5 times
that of normal nuclear matter~\cite{Danielewicz:2002pu}. Moreover,
low-density constraints on the equation of state of {\sl neutron-rich}
matter are starting to emerge from the distribution of fragments in
medium-energy collisions. In particular, by plotting the data in an
ingenious manner, a powerful scaling relation --- known as {\sl
isoscaling}~\cite{Tsang:2001jh,Tsang:2001dk} --- was uncovered and
shown to be sensitive to the low-density behavior of the symmetry
energy~\cite{Ono:2003zf}. Note that the symmetry energy equals to 
an excellent approximation the difference between the energy of 
pure neutron matter and that of symmetric matter.

Finally, enormous advances in both land- and spaced-based
observatories have brought the fields of nuclear physics and
astrophysics closer than ever before. In the particular case of
neutron star structure and its intimate connection to the equation of
state of dense matter, a few recent developments are worth
mentioning. Among these, observations of neutron-star--white-dwarf
binaries with the Arecibo radio telescope have resulted in the largest
neutron-star mass ever reported $M({\rm
PSR~J0751\!+\!1807})\!=\!2.1\!\pm\!0.2~M_{\odot}$~\cite{Nice:2005fi}.
The limiting mass of a neutron star represents the optimal (and
perhaps unique) way of constraining the high density component of the
equation of state. If the above error bars can be narrowed down any
further, a significant number of (soft) models of the equation of
state will be ruled out~\cite{Lattimer:2004pg,Lattimer:2006xb}.
Moreover, transient phenomena --- such as X-ray bursts powered by the
nuclear burning of H/He in low-mass X-ray binaries --- provide a
powerful observational constraint on the long-sought {\sl
mass-vs-radius} relationship of a neutron star. For example, the
mass-to-radius {\sl ratio} of the neutron star in the low-mass X-ray
binary EXO 0748-676 has been recently constrained by detecting the
gravitational redshift of certain absorption
lines~\cite{Cottam:2002cu}. More recently, by combining this
result with additional observational constraints, both the mass
{\sl and} radius of the neutron star in EXO 0748-676 were 
individually determined to be $M\!\ge\!2.10\!\pm\!0.28~M_{\odot}$ and
$R\!\ge\!13.8\!\pm\!1.80$~km~\cite{Ozel:2006bv} --- an interpretation
that appears to rule out {\sl all} soft equations of state.  For a
comprehensive recent account on constraints on the equation of state
from neutron-star observations see Ref.~\cite{Lattimer:2006xb}.

The manuscript has been organized as follows. In Sec.~\ref{Formalism}
a brief review of the relativistic formalism will be provided.
Particular emphasis will be placed on those model parameters that are
loosely constrained by existent ground-state observables and the
extent to which they may be firmly determined by the various
constraints discussed above. Next, in Sec.~\ref{Results} results will
be presented for a variety of low- and high-density observables that
will be compared against available theoretical, experimental, and
observational results. Finally, we offer a summary and conclusions
in Sec.~\ref{Conclusions}.

\section{Formalism}
\label{Formalism}

The Lagrangian density employed in this work is rooted on the seminal
work of Walecka, Serot, and their many collaborators (see
Refs.~\cite{Walecka:1974qa,Serot:1984ey,Serot:1997xg} and references
therein). Since first published by Walecka more than three decades
ago~\cite{Walecka:1974qa}, several refinements have been implemented
to improve the quantitative standing of the model. In the present work
we employ an interacting Lagrangian density of the following
form~\cite{Mueller:1996pm,Horowitz:2000xj,Todd-Rutel:2005fa}:
\begin{widetext}
\begin{align}
{\mathscr L}_{\rm int} & =
 \bar\psi\left[g_{\rm s}\phi   \!-\!
         \left(g_{\rm v}V_\mu  \!+\!
    \frac{g_{\rho}}{2}\tau\cdot{\bf b}_{\mu}
                               \!+\!
    \frac{e}{2}(1\!+\!\tau_{3})A_{\mu}\right)\gamma^{\mu}
         \right]\psi \nonumber \\
                   & -
    \frac{\kappa}{3!} (g_{\rm s}\phi)^3 \!-\!
    \frac{\lambda}{4!}(g_{\rm s}\phi)^4 \!+\!
    \frac{\zeta}{4!}
    \Big(g_{\rm v}^2 V_{\mu}V^\mu\Big)^2 \!+\!
    \Lambda_{\rm v}
    \Big(g_{\rho}^{2}\,{\bf b}_{\mu}\cdot{\bf b}^{\mu}\Big)
    \Big(g_{\rm v}^2V_{\mu}V^\mu\Big) \;.
 \label{Lagrangian}
\end{align}
\end{widetext}
The original Lagrangian density of Walecka consisted of an isodoublet
nucleon field ($\psi$) together with neutral scalar ($\phi$) and
vector ($V^{\mu}$) fields coupled to the scalar density
($\bar\psi\psi$) and conserved nucleon current
($\bar\psi\gamma^{\mu}\psi$), respectively~\cite{Walecka:1974qa}. In
spite of its simplicity (the model contains only two dimensionless
coupling constants), symmetric nuclear matter saturates even when 
the model was solved at the mean-field
level~\cite{Walecka:1974qa}. By adding additional contributions from a
single isovector meson ($b^{\mu}$) and the photon ($A^{\mu}$),
Horowitz and Serot~\cite{Horowitz:1981xw} obtained results for the
ground-state properties of finite nuclei that rivaled some of the most
sophisticated non-relativistic calculations of the time. However,
whereas the two dimensionless parameters in the original Walecka model
could be adjusted to reproduce the nuclear saturation point, the
incompressibility coefficient (now a prediction of the model) was too
large ($K\!\gtrsim\!500$~MeV) as compared with existing data on
breathing-mode energies~\cite{Youngblood:1977}. To overcome this
problem, Boguta and Bodmer introduced cubic ($\kappa$) and quartic
($\lambda$) scalar meson self-interactions that accounted for a
significant softening of the equation of state
($K\!=\!150\!\pm\!50$~MeV)~\cite{Boguta:1977xi}.  Two parameters of
the Lagrangian density of Eq.~(\ref{Lagrangian}) remain to be
discussed, namely, $\zeta$ and $\Lambda_{\rm v}$. Both of these
parameters are set to zero in the enormously successful NL3 model,
suggesting that the experimental data used in the calibration
procedure is insensitive to the physics encoded in these
parameters. Indeed, M\"uller and Serot found possible to build models
with different values of $\zeta$ that reproduce the same observed
properties at normal nuclear densities, but which yield maximum
neutron star masses that differ by almost one solar
mass~\cite{Mueller:1996pm}. This result indicates that observations of
massive neutron stars --- rather than laboratory experiments --- may
provide the only meaningful constraint on the high-density component
of the equation of state. Finally, the isoscalar-isovector coupling
constant $\Lambda_{\rm v}$ was added in Ref.~\cite{Horowitz:2000xj} to
modify the density dependence of the symmetry energy.  It was
found that models with different values of $\Lambda_{\rm v}$ reproduce
the same exact properties of symmetric nuclear matter, but yield
vastly different values for the neutron skin thickness of heavy nuclei
and for the radii of neutron stars~\cite{Horowitz:2001ya}.  The Parity
Radius Experiment (PREX) at the Jefferson Laboratory promises to
measure the skin thickness of $^{208}$Pb accurately and model
independently via parity-violating electron
scattering~\cite{Horowitz:1999fk, Michaels:2005}. PREX will provide a
unique experimental constraint on the density dependence of the
symmetry energy due its strong correlation to the neutron skin of
heavy nuclei~\cite{Brown:2000}.

\section{Results}
\label{Results}

In this section we compare the predictions from
NL3~\cite{Lalazissis:1996rd, Lalazissis:1999} and
FSUGold~\cite{Todd-Rutel:2005fa} against the theoretical,
experimental, and observational constraints discussed in the
Introduction.  The effective parameters of the two models are listed
in Table~\ref{Table1} and their predictions for several bulk
properties of nuclear matter are tabulated in Table~\ref{Table2}. Note
that $\rho_{0}$, $\varepsilon_{0}$, and $K$ denote the density, the
binding energy per nucleon and incompressibility coefficient of
symmetric nuclear matter while $J$ and $L$ denote the value and slope
of the symmetry energy, all at saturation density. In particular, the
pressure of pure neutron matter at saturation density is related to
$L$ through the following expression:
\begin{equation}
  P_{\rm PNM}(\rho_{0}) = \frac{1}{3}\rho_{0}L\;.
 \label{PPNM}
\end {equation} 

\begin{table}
\begin{tabular}{|l||c|c|c|c|c|c|c|c|}
 \hline
 Model & $m_{\rm s}$  & $g_{\rm s}^2$ & $g_{\rm v}^2$ & $g_{\rho}^2$
       & $\kappa$ & $\lambda$ & $\zeta$ & $\Lambda_{\rm v}$\\
 \hline
 \hline
 NL3     & 508.1940 & 104.3871  & 165.5854 &  79.6000
         & 3.8599 & $-$0.0159 & 0.0000 & 0.0000   \\
 \hline
 FSUGold & 491.5000 & 112.1996  & 204.5469 & 138.4701
         & 1.4203 & $+$0.0238 & 0.0600 & 0.0300   \\
\hline
\end{tabular}
\caption{Model parameters used in the calculations. The parameter
$\kappa$ and the inverse scalar range $m_{\rm s}$ are given in MeV.
The nucleon, omega, and rho masses are kept fixed at $M\!=\!939$~MeV,
$m_{\omega}\!=\!782.5$~MeV, and $m_{\rho}\!=\!763$~MeV, respectively.}
\label{Table1}
\end{table}

\begin{table}
\begin{tabular}{|l||c|c|c|c|c|}
 \hline
 Model & $\rho_{0}({\rm fm}^{-3})$ & $\varepsilon_{0}$(MeV) 
       & $K$(MeV) & $J$(MeV) & $L$(MeV) \\ 
 \hline
 \hline
 NL3     & 0.148 & -16.24 & 271 & 37.3 & 118.4 \\
 FSUGold & 0.148 & -16.30 & 230 & 32.6 &  60.5 \\
\hline
\end{tabular}
\caption{Bulk parameters characterizing the energy of symmetric 
         nuclear matter ($\rho_{0}$, $\varepsilon_{0}$, and $K$) 
         and the symmetry energy ($J$ and $L$) at saturation 
         density.} 
\label{Table2}
\end{table}

\subsection{Theoretical Constraints}
\label{Theory}

Perhaps surprisingly, one of the most stringent constraints on the
equation of state of low density neutron-rich matter emerges from
theoretical considerations, namely, from the universality of dilute
Fermi gases with an ``infinite'' scattering length ($a$).  In this
limit the only energy scale in the problem is the Fermi energy
($\varepsilon_{\rm F}$), so the energy per particle is constrained to
be that of the free Fermi gas up to a dimensionless {\sl universal
constant} ($\xi$) that is independent of the details of the two-body
interaction~\cite{Carlson:2003}. That is,
\begin{equation}
  \frac{E}{N} = \xi\frac{3}{5}\varepsilon_{\rm F} \;.
\end{equation}
To date, the best theoretical estimates place the value of the
universal constant around $\xi\!\approx\!0.4$~\cite{Baker:1999dg,
Heiselberg:2000bm,Carlson:2003,Nishida:2006br}.

\begin{figure}[ht]
\vspace{0.50in}
\includegraphics[width=5in,angle=0]{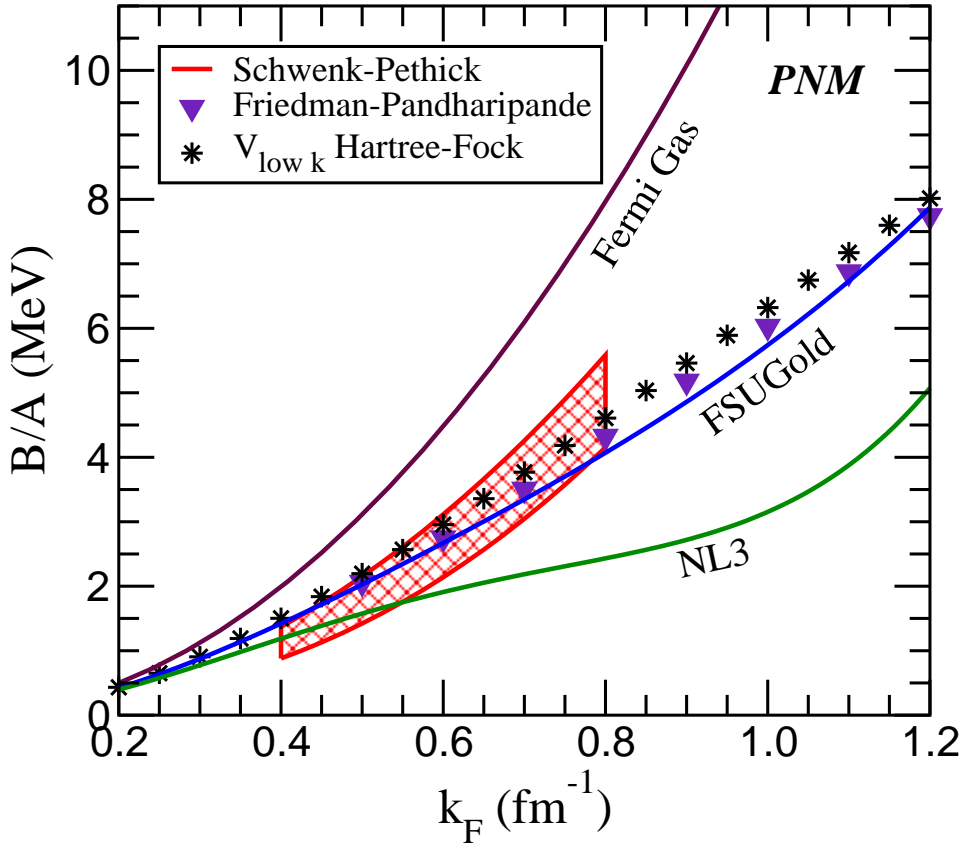}
\caption{(color online) Equation of state of pure neutron matter
         as a function of the Fermi momentum. Predictions  are 
         shown for the accurately calibrated 
         NL3~\cite{Lalazissis:1996rd,Lalazissis:1999} (green
         line) and  FSUGold~\cite{Todd-Rutel:2005fa} (blue line) 
         parameter sets. Shown also are various microscopic 
         descriptions --- including a {\sl model-independent} 
         result based on the physics of resonant Fermi gases by 
         Schwenk and Pethick~\cite{Schwenk:2005ka} (red region).} 
\label{Fig1}
\end{figure}

While the neutron-neutron scattering length is large indeed 
($a_{\rm nn}\!=\!-18.5$~fm), pure neutron matter deviates from 
unitarity due to a non-negligible value of the effective range 
of the neutron-neutron interaction ($r_{\rm e}\!=\!+2.7$~fm).
Thus, corrections to the low-density equation of state of pure neutron
matter must be computed for $k_{\rm F}\!\sim\!r_{\rm
e}^{-1}\simeq0.4~{\rm fm}^{-1}$. Such corrections have been recently
computed by Schwenk and Pethick~\cite{Schwenk:2005ka}, with their
results displayed as the red region in Fig~\ref{Fig1}. Also shown are
the predictions of two microscopic models based on realistic two-body
interactions, one of them being the venerated equation of state of
Friedman and Pandharipande~\cite{Friedman:1981qw}. Finally, the
predictions of NL3 and FSUGold are also shown.  It is gratifying that
the softening of the symmetry energy of FSUGold --- caused by
incorporating constraints from breathing-mode
energies~\cite{Todd-Rutel:2005fa} --- appears consistent with the
physics of resonant Fermi gases.  Such a powerful universal constraint
should be routinely and explicitly incorporated into future
determinations of density functionals. Indeed, such a constrain 
appears to rule out many of the models displayed in Fig.~2 of
Ref.~\cite{Brown:2000}.

\subsection{Experimental Constraints}
\label{Experiment}

\begin{figure}[ht]
\vspace{0.50in}
\includegraphics[width=5in,angle=0]{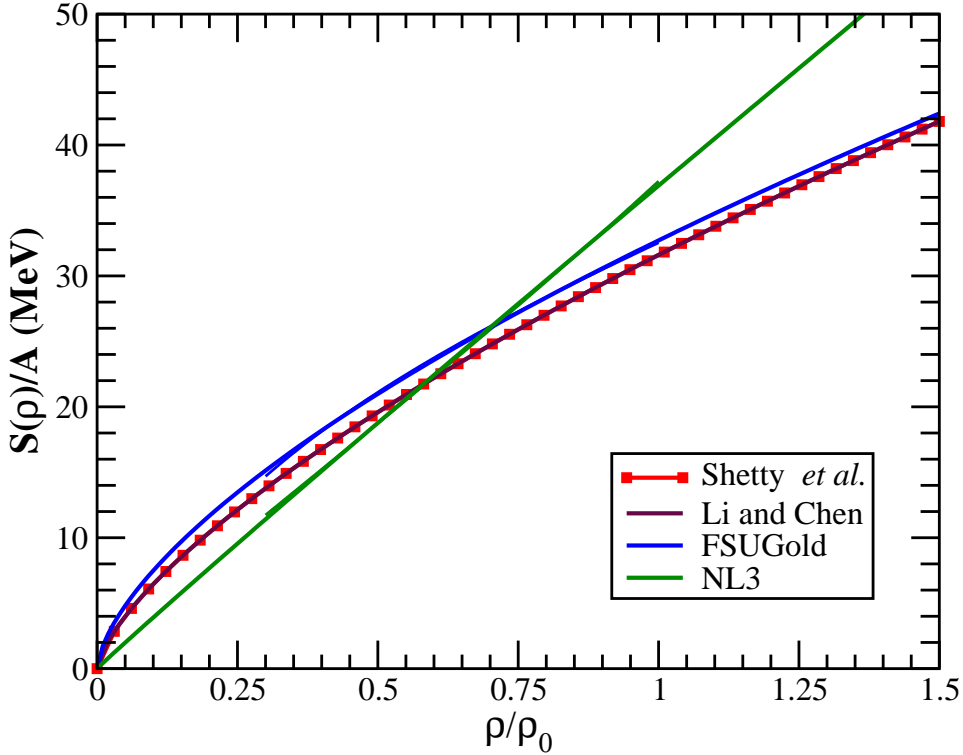}
\caption{(color online) Symmetry energy as a function of the
         baryon density (expressed in units of the saturation 
         density $\rho_{0}\!=\!0.148~{\rm fm}^{-3}$). Predictions  
         are shown from the NL3~\cite{Lalazissis:1996rd,
         Lalazissis:1999} (green line) and  
         FSUGold~\cite{Todd-Rutel:2005fa} (blue line) models. 
         Shown also are the results from Li and
         Chen~\cite{Li:2005jy} (maroon line) and the experimental 
         analysis of Shetty and collaborators~\cite{Shetty:2005qp}
         (red symbols).}
\label{Fig2}
\end{figure}

Laboratory experiments place important constraints on the equation of
state of hadronic matter. Indeed, a variety of ground-state properties
(primarily masses and charge radii) of finite nuclei are routinely
incorporated into the calibration procedure of the models. However,
the impact of heavy-ion experiments on these models is just starting
to emerge. A particularly relevant example involves the distribution
of fragments in medium-energy heavy ion collisions. It has been shown
that the {\sl ratio} of isotopic yields [$R_{21}(N,Z)$] obeys a
powerful scaling relation~\cite{Tsang:2001jh,Tsang:2001dk} that is
sensitive to the low-density behavior of the symmetry
energy~\cite{Ono:2003zf}.  It is observed, quite naturally, that the
reaction with neutron-rich nuclei ({\it e.g.,}
${}^{124}$Sn+${}^{124}$Sn) produces more neutron-rich and less
proton-rich fragments relative to the neutron-deficient reaction ({\it
e.g.,} ${}^{112}$Sn+${}^{112}$Sn). This makes the reaction yields
particularly sensitive to the density dependence of the symmetry
energy.  For example, a stiff symmetry energy (such as the one
displayed by NL3 in Fig.~\ref{Fig2}) imposes a stiff penalty on the
system at high density for departing from the symmetric ($N\!=\!Z$)
limit. It is, however, the softer symmetry energy (such as FSUGold in
Fig.~\ref{Fig2}) that imposes the stiffer penalty at the {\sl low
densities} relevant to the multifragmentation process. As such, one
expects more neutron-rich fragments to be produced by a stiff rather
than by a soft equation of state~\cite{Liu:2002tp}. The experimental
signature of this behavior is imprinted in a parameter than controls 
the variation of $R_{21}(N,Z)$ with $N$ for fixed $Z$ (a parameter 
usually denoted by $\alpha$). Using general thermodynamic arguments, 
the value of $\alpha$ was related to the change of the neutron 
chemical potential with the neutron-proton 
asymmetry~\cite{Tsang:2001jh,Tsang:2001dk}. Note that it was
precisely the study of the neutron chemical potential of 
neutron-rich nuclei at zero temperature ({\sl i.e.,} the neutron 
Fermi energy) that lead to the conclusion that models with a 
soft symmetry energy reach the neutron-drip line before those 
with a stiffer symmetry energy~\cite{Todd:2003xs}.

In a recent study, Shetty, Yennello, and Souliotis~\cite{Shetty:2005qp} 
used the scaling behavior of the fragment yields --- coupled to a 
molecular dynamics simulation --- to extract the density dependence of 
the symmetry energy. In the low density regime probed in the collisions, 
the density dependence of the symmetry energy may be parametrized according 
to the following simple formula:
\begin{equation}
 S(\rho)=S(\rho_{0})\left(\frac{\rho}{\rho_{0}}\right)^{\gamma}\;.
 \label{SymmEnergy}
\end{equation}
To make contact with this approach, our theoretical results were
fitted to the above formula in the $\rho\!=\!(0.3\!-\!1.0)\rho_{0}$
range, yielding values for the symmetry energy at saturation
[$S(\rho_{0})$] and for the exponent ($\gamma$) as displayed in
Eq.~(\ref{SymmEParams}). This same information is depicted in
graphical form in Fig.~\ref{Fig2}. As in the case of pure neutron
matter (see Fig.~\ref{Fig1}) it appears that the density dependence of
the symmetry energy predicted by FSUGold --- at least at low densities
--- is consistent with this experimental analysis.  However, the
experimental determination is not without
controversy~\cite{Ono:2005vv,Shetty:2006hh}. Moreover, one should be
aware that the connection between the collision of heavy ions and the
{\sl zero-temperature} equation of state is model dependent.
\begin{equation}
 S(\rho_{0})=
  \begin{cases} 
     31.6~{\rm MeV}, & \text{Ref.~\cite{Shetty:2005qp},}  \\
     31.6~{\rm MeV}, & \text{Ref.~\cite{Li:2005jy},}  \\
     32.7~{\rm MeV}, & \text{FSUGold,} \\
     36.9~{\rm MeV}, & \text{NL3;} 
 \end{cases} \qquad
 \gamma=
  \begin{cases} 
     0.69, & \text{Ref.~\cite{Shetty:2005qp},}  \\
     0.69, & \text{Ref.~\cite{Li:2005jy},}  \\
     0.64, & \text{FSUGold,} \\
     0.98, & \text{NL3.} 
 \end{cases} 
 \label{SymmEParams}
\end{equation}

\begin{figure}[ht]
\vspace{0.50in}
\includegraphics[width=5in,angle=0]{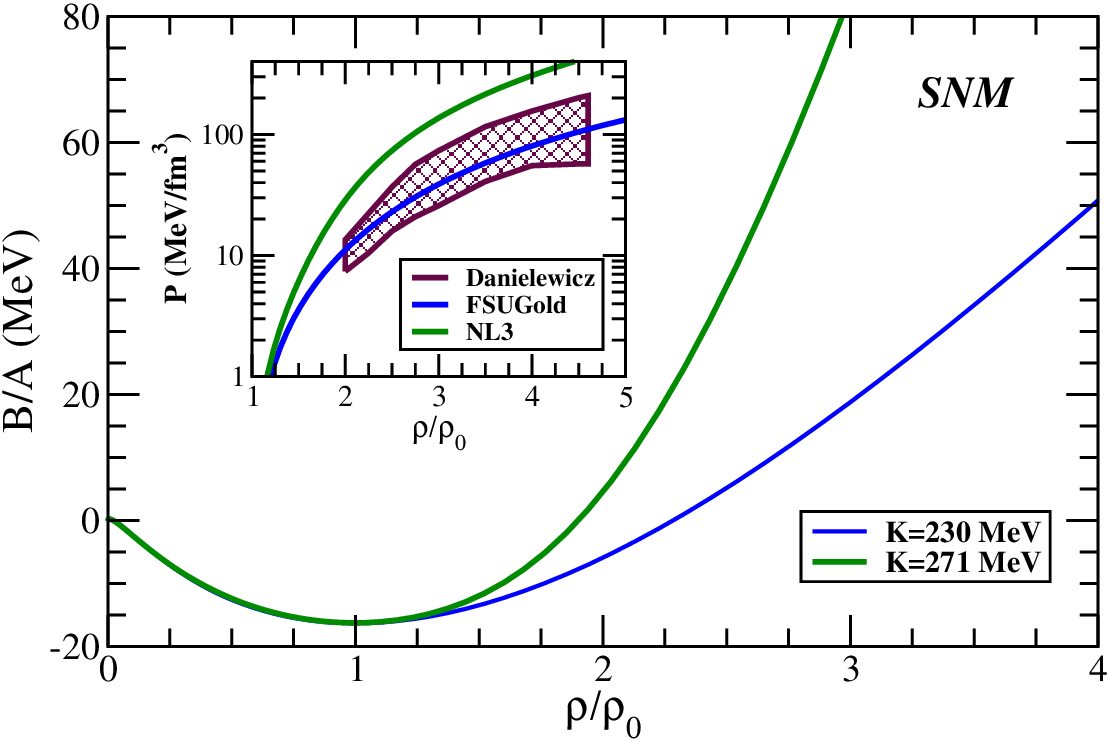}
\caption{(color online) Binding energy per nucleon as a function
         of baryon density (expressed in units of the saturation 
         density $\rho_{0}\!=\!0.148~{\rm fm}^{-3}$) for symmetric 
         nuclear matter. Theoretical predictions are shown for the  
         NL3~\cite{Lalazissis:1996rd,Lalazissis:1999} (green line)
         and FSUGold~\cite{Todd-Rutel:2005fa} (blue line) models.
         Shown in the inset is a comparison between the equation 
         of state extracted from energetic nuclear 
         collisions~\cite{Danielewicz:2002pu} and the predictions 
         of these two models.} 
\label{Fig3}
\end{figure}

\begin{figure}[ht]
\vspace{0.60in}
\includegraphics[width=5in,angle=0]{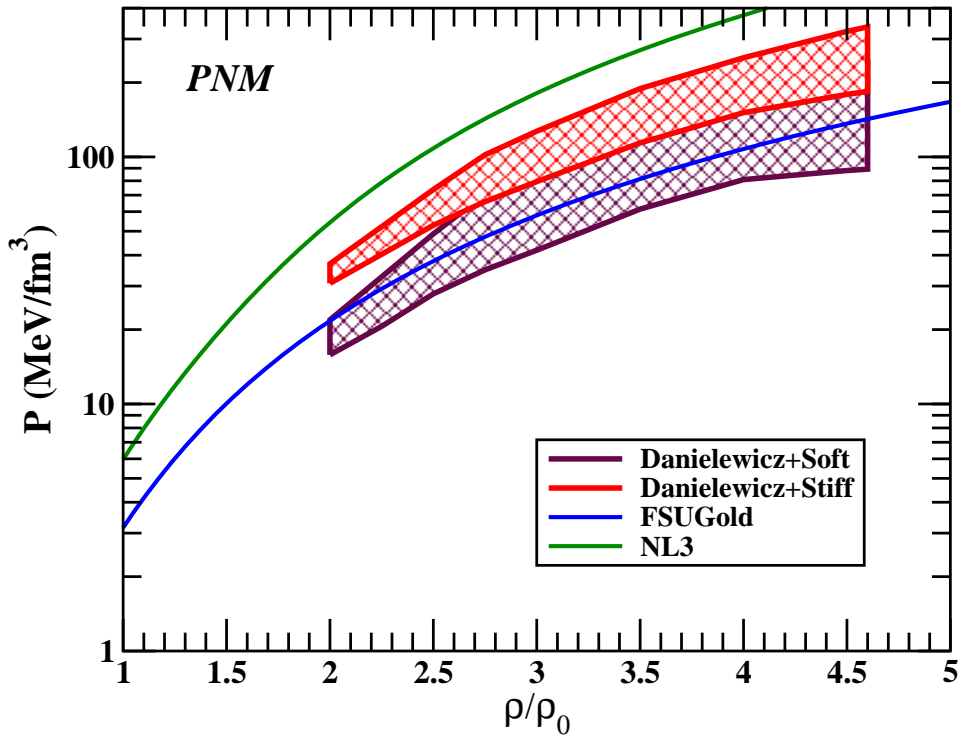}
\caption{(color online) The pressure as a function of baryon density 
         (expressed in units of the saturation density 
         $\rho_{0}\!=\!0.148~{\rm fm}^{-3}$) for pure neutron matter 
         as extracted from energetic nuclear 
         collisions~\cite{Danielewicz:2002pu} by assuming soft
         (maroon) and stiff (red) symmetry energies. Also shown
         are the theoretical predictions from the
         NL3~\cite{Lalazissis:1996rd,Lalazissis:1999} (green line)
         and FSUGold~\cite{Todd-Rutel:2005fa} (blue line) models.}
\label{Fig4}
\end{figure}

Nuclear collisions may also be used to constrain the high-density
behavior of nucleonic matter. To illustrate this point we display in
Fig.~\ref{Fig3} the binding energy per nucleon of {\sl symmetric}
nuclear matter as a function of the baryon density as predicted by
both the NL3 and FSUGold models. Note that both models reproduce the
equilibrium properties of symmetric nuclear matter and display the
same {\sl quantitative} behavior at densities below the saturation
point.  Yet their high-density predictions are significantly
different.  This emerges from a combination of two factors. First,
FSUGold predicts an incompressibility coefficient $K$ considerably
lower than NL3, namely, $230$~MeV {\sl vs} $271$~MeV (see
Table~\ref{Table2}). Second, and more importantly, FSUGold 
includes a self-energy coupling
[denoted by $\zeta$ in Eq.~(\ref{Lagrangian})] that is responsible for
a significant softening at high density. Note that the {\sl mixed}
isoscalar-isovector coupling ($\Lambda_{\rm v}$) plays no role in
symmetric nuclear matter. We now compare the predictions of these
two models against results obtained from energetic nuclear collisions
that can compress baryonic matter to densities as high as those
predicted to exist in the core of neutron stars. The inset in
Fig.~\ref{Fig3} provides us with such a comparison. By analyzing the
manner in which matter flows after the collision of two energetic gold
nuclei, the equation of state of {\sl symmetric} nuclear matter was
extracted up to densities of 4-to-5 times saturation
density~\cite{Danielewicz:2002pu}. Figure~\ref{Fig3} seems to rule out
overly stiff equations of state (such as NL3). And while it continues
to be gratifying that FSUGold is consistent with this analysis, one
should reiterate that the connection between energetic nuclear
collisions and the equation of state of cold nuclear matter is model
dependent. Yet within these limitations, the same
analysis~\cite{Danielewicz:2002pu} has been used to impose constraints
on the equation of state of {\sl pure neutron matter} by assuming two
models --- one soft and one stiff --- for the unknown density
dependence of the symmetry energy. The resulting equations of state 
are displayed in Fig.~\ref{Fig4} alongside the predictions of both
theoretical models.  The FSUGold parametrization appears consistent
with a relatively soft symmetry energy.

\subsection{Observational Constraints}
\label{Observational}
Ultimately, the most reliable constraints on the high-density
component of the equation of state will come from astronomical
observations. Indeed, important constraints are starting to emerge
from the combination of a large number of
observations~\cite{Lattimer:2006xb}. Here we limit ourselves to only
two of them for their significant impact on the present analysis.
For two other studies similar in spirit to the present one see
the very recent references~\cite{Sagert:2007nt} 
and~\cite{Blaschke:2007fr}.

\begin{figure}[ht]
\vspace{0.50in}
\includegraphics[width=5in,angle=0]{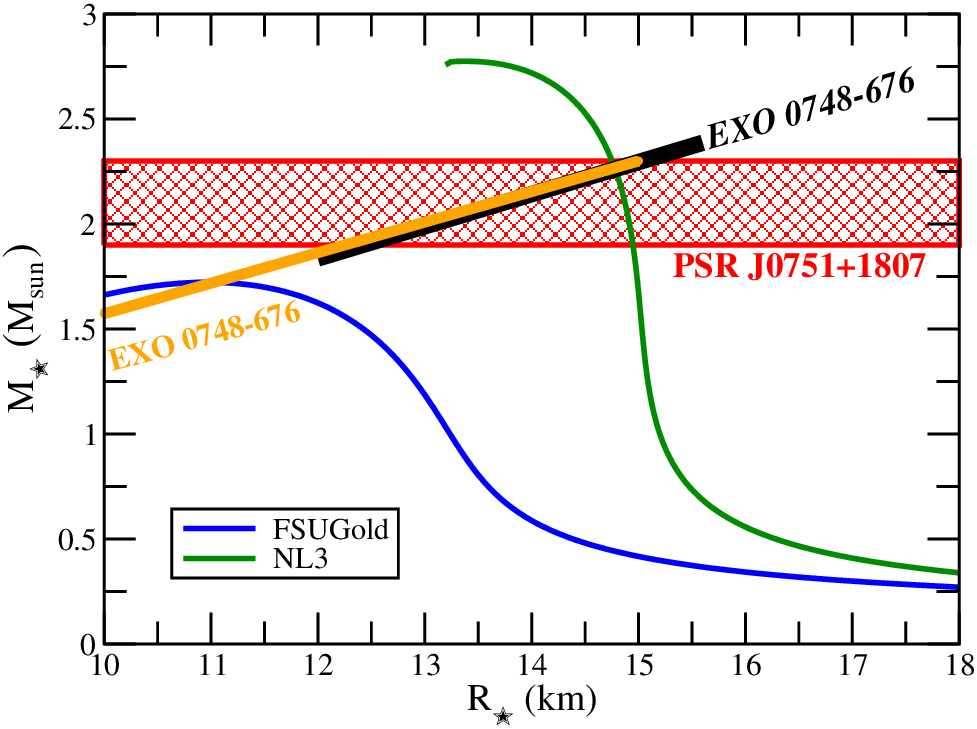}
\caption{(color online) Constraints on the mass-vs-radius
         relationship of neutron stars. Displayed in red is
         the allowed region as determined by the analysis of
         Nice and collaborators~\cite{Nice:2005fi}. The
         black and orange solid lines result from the analyzes 
         of EXO 0748-676 by \"Ozel~\cite{Ozel:2006bv}, and 
         Villarreal and Strohmayer~\cite{Villarreal:2004nj}, 
         respectively. Also shown are the theoretical predictions 
         from the NL3~\cite{Lalazissis:1996rd,Lalazissis:1999} 
         (green line) and FSUGold~\cite{Todd-Rutel:2005fa} 
         (blue line) models.}
\label{Fig5}
\end{figure}

The first observation that impacts significantly on the present 
work is the one by Nice and collaborators at the Arecibo radio
telescope~\cite{Nice:2005fi}. Such observation of a
neutron-star--white-dwarf binary system appears to suggest a
neutron-star mass of 
$M({\rm PSR~J0751\!+\!1807})\!=\!2.1\!\pm\!0.2~M_{\odot}$ (this 
is denoted by the red region in Figure~\ref{Fig5}). This appears 
to be the largest neutron-star mass ever reported and one that
is significantly larger than those most accurately determined from 
double neutron-star binaries that display a mean of only about
1.35-1.40~$M_{\odot}$ with a very small dispersion. If the limits on
the mass of PSR~J0751\!+\!1807 can be tighten any further (after 
all, at the $2\sigma$ level the observation accommodates the rather 
wide range of 1.6-2.5~$M_{\odot}$) it would practically pin down the
high-density component of the equation of state. Indeed, in the
particular case of the FSUGold model with a prediction of only
1.72~$M_{\odot}$ for the limiting mass, this observation would 
demand a mild hardening of the equation of state at high densities. 
As first observed by M\"uller and Serot~\cite{Mueller:1996pm}, in 
relativistic mean-field models this may be efficiently achieved 
by simply tuning the non-linear coupling $\zeta$, while leaving
intact all other parameters of the model.

The second observation that seems to suggest a hard equation of state
is that of the low-mass X-ray binary EXO 0748-676. The first
constraint on the equation of state from such an object came from the
detection of gravitationally redshifted absorption lines in Oxygen and
Iron by Cottam and collaborators~\cite{Cottam:2002cu}. By measuring a
gravitational redshift of $z\!=\!0.35$, the mass-to-radius {\sl ratio}
of the neutron star gets fixed at $M/R\!\simeq\!0.15$ (with $M$
expressed in solar masses and $R$ in kilometers). By incorporating
additional constraints arising from Eddington and thermal fluxes, a
recent analysis by \"Ozel seems to place {\sl simultaneous} limits on
the mass and radius of the neutron star in EXO 0748-676. That is,
$M\!\ge\!2.10\!\pm\!0.28~M_{\odot}$ and
$R\!\ge\!13.8\!\pm\!1.80$~km~\cite{Ozel:2006bv}. These limits are
indicated by the black solid line in Fig.~\ref{Fig5}. An earlier
determination of the spin frequency of the same neutron star by
Villarreal and Strohmayer~\cite{Villarreal:2004nj}, when combined with
the rotational broadening of surface spectral lines, yields an
independent determination of the stellar radius of
$R\!\approx\!11.5^{+3.5}_{-2.5}$~km. This estimate, when combined with
the gravitational redshift, yields the orange line in
Fig.~\ref{Fig5}. Note, however, that the use of rotational broadening
to constrain the stellar radius has been put into question in
Ref.~\cite{Ozel:2006bv}. Finally, {\sl mass-vs-radius} predictions
from the NL3 and FSUGold models are displayed in Fig.~\ref{Fig5}.  The
results clearly indicate the significantly harder character of the
equation of state predicted by NL3 relative to FSUGold. This, even
when both models predict practically identical properties for existent
ground-state observables of finite nuclei. Based solely on these
observations, NL3 with its stiff equation of state appears to fair far
better than FSUGold, despite the fact that both breathing-mode
energies and heavy-ion experiments seem to suggest that the NL3
equation of state is overly stiff.

\section{Conclusions}
\label{Conclusions}

Accurately calibrated relativistic models of nuclear structure have
been enormously successful at describing a variety of ground-state
properties throughout the periodic table by employing a relatively
small number of effective parameters.  Chief among these is the NL3
parameter set of Lalazissis, Ring, and
collaborators~\cite{Lalazissis:1996rd,Lalazissis:1999}.  Yet under
closer scrutiny, it was revealed, perhaps not surprisingly, that the
properties employed in the calibration procedure of such models are
insufficient to firmly pin down the equation of state even around 
saturation density. In an effort to lift this {\sl ``degeneracy''} 
the FSUGold model was conceived~\cite{Todd-Rutel:2005fa}. Relative 
to NL3, FSUGold includes two additional effective parameters. 
One of them [denoted by $\Lambda_{\rm v}$ in Eq.~(\ref{Lagrangian})] 
is responsible for a softening of the symmetry energy and should be 
firmly constrained by the pioneering {\sl Parity Radius Experiment} 
at the Jefferson Laboratory that is scheduled to run in the early 
part of 2009~\cite{Michaels:2005}. The other parameter [denoted by 
$\zeta$ in Eq.~(\ref{Lagrangian})] controls the high-density 
component of the equation of state and may be tuned to generate 
maximum neutron star masses that differ by almost one solar mass
without significantly affecting the behavior of the equation of
state near saturation density ~\cite{Mueller:1996pm}. These two 
parameters --- that are set to zero in the NL3 model --- have a 
particularly dramatic impact on two observables: (a) the neutron-skin
thickness of ${}^{208}$Pb and (b) the limiting mass of a neutron star.
In both cases NL3 predicts significantly larger values than FSUGold,
namely, 0.28~fm {\sl vs} 0.21~fm for the former and 2.78~$M_{\odot}$
{\sl vs} 1.72~$M_{\odot}$ for the latter. The main goal of this
contribution was to test the validity of these models away from
their region of applicability. 

The validation of FSUGold --- a model calibrated to various
ground-state properties and collective excitations of finite nuclei
--- was implemented through a detailed comparison against recent
theoretical, experimental, and observational constraints. These
constraints emerged from the universal behavior of dilute Fermi gases
with large scattering lengths~\cite{Schwenk:2005ka}, heavy-ion
experiments that probe both the low- and high-density domain of the
equation of state~\cite{Tsang:2001jh,Tsang:2001dk,Shetty:2005qp,
Danielewicz:2002pu}, and astronomical observations that place limits
on masses and radii of neutron stars~\cite{Nice:2005fi, Ozel:2006bv}.
On the basis of these comparisons, it was concluded that FSUGold meets
all the challenges, even when no attempt was ever made to incorporate
these constraints into the calibration procedure. If at all, only the
observational data seems to call FSUGold into question by suggesting a
slightly harder equation of state. The promise of improved
observations and analyzes with existent and future missions, such as
{\sl Constellation X}, offers the greatest hope for determining the
high-density component of the equation of state.

The response to such future achievements should not be limited to an
indiscriminate adjustment of parameters. After all, {\sl ``knobs''}
can always be turned to reproduce a particular set of experiments
and/or observations. Rather, models should aim at reproducing {\sl
simultaneously} a myriad of observables that probe the equation of
state over a wide dynamic range. This should be one of the primary
missions of all modern theoretical approaches. Only by adopting such
strict standards one could test the validity and applicability of such
models. Only then can one ensure the emergence of exotic phenomena.

\begin{acknowledgments}
The author is grateful to the Institute for Nuclear Theory at the
University of Washington for its hospitality during the early phase of
this work. The author is also grateful to Achim Schwenk for providing
access to his theoretical results. This work was supported in part by
the Department of Energy grant DE-FD05-92ER40750.
\end{acknowledgments}

\vfill\eject
\bibliography{ReferencesJP}

\end{document}